\RequirePackage{fix-cm}
\documentclass[smallextended]{svjour3}       
\smartqed  
\usepackage{graphicx}
\begin{document}

\title{Simulations of orbital debris clouds due to breakup events and their characterisation using the Murchison Widefield Array radio telescope
}

\titlerunning{Debris cloud characterisation with the MWA}        

\author{Wynand Joubert \and Steven Tingay
}


\institute{W. Joubert \at
              International Centre for Radio Astronomy Research, Curtin University, Bentley, WA 6102, Australia \\
              \email{wynand.joubert@student.curtin.edu.au}           
           \and
           S. Tingay \at
              International Centre for Radio Astronomy Research, Curtin University, Bentley, WA 6102, Australia
}

\date{Received: date / Accepted: date}

\maketitle

\begin{abstract}
In this paper we consider the use of wide field of view radar sensors such as the Murchison Widefield Array (MWA), a low frequency radio telescope designed for astrophysics and cosmology, for rapid response observations of the debris clouds produced by collisions between objects in Earth orbit.  With an increasing density of objects in Low Earth Orbit, including legacy assets used by the astronomy community over decades, the risk of new debris clouds forming is also increasing.  The MWA constitutes a wide field, rapid response passive radar system and we explore its likely performance in the detection and characterisation of debris clouds.  In general, astronomy facilities such as the MWA can play a role in protecting the space environment for the future.  In order to undertake this work, we adapt the NASA EVOLVE 4.0 breakup model, utilising the EVOLVE outputs to produce representative dynamic debris clouds.  We find that the MWA is likely to detect a large fraction ($>70\%)$ of modelled debris cloud fragments for collision masses between 100 kg and 1000 kg for orbits in the lower part of LEO, if the MWA can achieve close to optimal detection sensitivity.  Useful detection fractions are still achieved for more conservative assumptions.  The detection fraction of fragments decreases as a function of altitude and inversely with collision mass.  Encouragingly, we find that the wide field nature of the MWA allows the full evolving debris clouds to be observed in a single observation, with only $\sim2\%$ of the debris fragments escaping the sensitive portion of the field of view after 100 seconds, for all collision masses and altitudes.  These results show that the MWA is an intrinsically useful facility for the rapid characterisation of debris clouds, but that work is required to achieve the data processing within an appropriate timeframe to provide rapid alerts.
\end{abstract}

\section{Introduction}
\label{intro}

With each object that is launched into orbit, the potential for collisions between them increases. The United Nations Office for Outer Space Affairs has maintained an active index of objects launched into outer space\footnote{http://www.unoosa.org/oosa/en/spaceobjectregister/index.html}.  As of March 2020, 9,286 objects have been launched into Earth orbit, a 21\% increase since the end of 2015. These objects occupy a region of space ranging from 250 to 40,000 kilometres above Earth, where 62.9\% reside in Low Earth Orbit (LEO; below 2,000 km), a region where orbital speeds are around 8 km/s. 

At these speeds, collisions are catastrophic in nature, as witnessed in 2009 in an unintentional collision between Iridium 33 and Cosmos 2251 at an altitude of 785 km \cite{ODQN}.  The satellites collided at a near right angle at a relative speed of 11,700 m/s, which resulted in the creation of two thousand catalogued fragments, according to the US Space Surveillance Network (SSN) \cite{SSN} by 2011 with approximate cross sections of ten centimetres. One such fragment was projected to come into close proximity of the International Space Station in March of 2011, which called for a full evacuation. Despite small sizes, such fragments can inflict substantial damage due to their high speeds. Detections of numerous smaller fragments from this collision with cross sections as small as a centimetre were obtained by the sensitive Haystack Auxiliary and Goldstone radars \cite{HG}. These fragments can have lifetimes measured in decades, allowing for greater accumulation and further risk of the Kessler Syndrome \cite{1978JGR....83.2637K}, a runaway cascade of collisions that renders entire orbits unusable.

Analyzing the characteristics and behaviour of debris clouds over time provides further insight into the scope of the problem. In January of 2007, the Fengyun-1C spacecraft was intentionally destroyed in a collision with a ballistic kinetic kill vehicle (KKV). This resulted in one of the most substantial debris clouds ever observed, with more than two thousand fragments detected shortly after the collision with an additional estimated 150,000 fragments created \cite{fengyun,2007amos.confE..35K}. As of May 2019, the CelesTrak satellite catalogue\footnote{https://celestrak.com/} has 3,442 recorded pieces of debris from this event, of which only 621 have de-orbited in twelve years.  Other such tests have been conducted by Russia, the US, and recently by India \cite{akhmetov2019analysis}, in each case creating debris clouds\footnote{https://en.wikipedia.org/wiki/Anti-satellite\_weapon}.

In addition to the threat to assets in space, activities in space and ground-based astronomy are starting to come into conflict, by virtue of the rise of large constellations of satellites \cite{2020arXiv200110952G}, the primary upcoming cause of congestion in space.  However, it is also the case that the astronomy community has a significant number of legacy assets in space that are potential participants in the problem.  For example, on January 29, 2020, the defunct infrared telescope IRAS (Infrared Astronomical Satellite) \cite{1984ApJ...278L...1N} had a near miss with GGSE 4 (Gravity Gradient Stabilization Experiment 4).  At a relative speed in excess of 14 km/s and with IRAS a 1,100 kg mass, the two satellites passed within an estimated 13 - 87 m of each other \footnote{https://en.wikipedia.org/wiki/GGSE-4}.  

IRAS, a highly productive product of the astronomy community, therefore now poses a collision hazard in space.  However, the astronomy community has an ability to contribute to Space Situational Awareness (SSA) activities, to assist in characterizing and mitigating against these hazards.  Many ground-based telescopes, optical and radio, can be utilised for SSA, for example in  \cite{2020AcAau.167..374L,2019AdSpR..64.1527L,2019amos.confE..41R}.  While the individual utility of a single ground-based telescope for SSA is limited, the global network of astronomical facilities is formidable, including some unique capabilities, as described below for the Murchison Widefield Array radio telescope in relation to on-orbit collisions.

Given the increasing risk of collisions in orbit, and the increased probability that debris clouds will result, an emerging challenge is the rapid detection and characterisation of debris clouds and the prediction of their evolution with time.  For objects in LEO, since a debris cloud of thousands of objects will rapidly expand, as well as traverse a full orbit in $\sim$90 minutes, rapid observations of the full debris cloud are required in order to quickly characterise the orbits for as many debris fragments as possible.  Thus, wide field of view sensors are required.  One such sensor, considered in this paper, is the Murchison Widefield Array (MWA), a low frequency (70 - 300 MHz) interferometric radio telescope located in Western Australia \cite{2013PASA...30....7T}.  The MWA was built for astrophysics and cosmology, but has also been shown to constitute a very effective passive radar system for debris in space, utilising FM radio transmitters as illuminators of opportunity.  Passive radar techniques using the MWA have been demonstrated by \cite{2013AJ....146..103T,7944483,8835821,2020arXiv200201674P}.  One of the defining characteristics of the MWA is its field of view, at FM frequencies spanning in excess of a thousand square degrees at high sensitivity, and the full accessible celestial hemisphere at reduced sensitivity.  Another defining feature is the MWA's full electronic directional steering, allowing rapid response re-pointing.  Thus, the MWA is well placed to react to the creation of new debris clouds due to collisions and to observe the debris clouds in their entirety utilising passive radar techniques.

In this paper, we explore the likely performance of the MWA in this scenario.  We utilise existing models for the creation of debris clouds.  There are many highly complex factors that govern the number and distribution of fragments that result from a collision in orbit. This creates very substantial challenges when attempting to model and predict the result of any given collision. The variation in shape, mass distribution, and use of materials across all satellite configurations cannot be easily accounted for when attempting theoretical collision modelling. The difficulties are further compounded by the fact that these collisions are classified as hypervelocity collisions (collision speed greater than roughly 3000 m/s) where the metals and materials behave as fluids due to the overwhelming inertial stresses.

From as early as the 1970’s, NASA has been developing a breakup model called EVOLVE \cite{2001AdSpR..28.1377J}. EVOLVE takes an empirical approach, constraining a parameterised model based on data collected for numerous incidents involving both satellites and rocket bodies. Given a set of input parameters, such as the mass of the colliding system, EVOLVE generates distributions for the number of fragments as a function of size (and therefore mass) and ejection speed from the collision.  These models are parameterised differently for collisions and explosions.

We use EVOLVE as a basis for our collision simulations (but we note that we can also utilise our simulations to characterise explosions as well as collisions, although we restrict ourselves to collisions in this paper).  However, some adaptation is required for our purposes.  EVOLVE does not track the dynamics of a collision, in order to generate a debris cloud in which each fragment has an ejection speed and direction.  Nor does EVOLVE impose physical constraints such as the conservation of momentum or mass on collisions.  In order to generate realistic and representative debris cloud simulations, we use the EVOLVE outputs and add approximate collision dynamics.  This allows us to then examine the MWA's response to an evolving debris cloud using different collision masses, fragment sizes, collision altitudes etc.

In Section 2, we describe our adaptation of EVOLVE in order to run debris cloud simulations.  In Section 3, we present the results of running a set of simulations and the likely performance of the MWA in detecting and characterising the debris clouds.  In Section 4, we discuss our results and conclusions from this work.

\section{Adaptation of the NASA Breakup Model}
\label{model}
In this section, we describe the debris cloud simulations we have developed, based on the NASA EVOLVE 4.0 breakup model \cite{2001AdSpR..28.1377J}.  We have implemented this code ourselves, as we were unable to source any code implemented by \cite{2001AdSpR..28.1377J}.  The code we have implemented for this work is freely available via a GitHub respository\footnote{https://github.com/steven-tingay/Orbital-Collisions}.

In this work, we emphasise that we are not seeking to develop an accurate or comprehensive simulation of any particular collision event. Rather, we aim to generate physically plausible representative collision simulations that broadly generate a realistic number of fragments, with a realistic distribution of sizes, and which produce realistically evolving debris clouds.

First, we provide a short summary of EVOLVE 4.0, describing our code implementation of the model and verification against results published by \cite{2001AdSpR..28.1377J}.  We then describe our use of the EVOLVE 4.0 outputs to generate the dynamics of realistic debris clouds.  EVOLVE 4.0 deals with on-orbit explosions and collisions (catastrophic and non-catastrophic).  We limit ourselves to the case of catastrophic collisions in this paper (relative kinetic energy of the smaller object divided by the mass of the larger object is equal to or greater than 40 J/g).

EVOLVE 4.0 utilises a quantity to represent fragment size resulting from a collision, the characteristic length, $L_{c}$ (units of m).  Based on empirical data, EVOLVE 4.0 uses a powerlaw distribution for the number of fragments generated larger than $L_{c}$, that is also dependent on the summed masses of the colliding objects, $M$ (units of kg; in the remainder of the paper, $M$ is referred to as the collision mass).  This characteristic length distribution drives the model.

Also based on empirical data, EVOLVE 4.0 maps characteristic length to fragment area-to-mass ratio, via a set of distributions.  Thus, a given $L_{c}$ corresponds to a particular area-to-mass distribution that should be randomly sampled $n_{L_{c}}$ times, where $n_{L_{c}}$ is the number of fragments of characteristic length $L_{c}$.  In this manner, every fragment generated by the model is assigned an area-to-mass ratio, which can be converted to a mass via an expression that relates area to $L_{c}$.

EVOLVE 4.0 takes a similar approach to generating ejection speeds for the fragments, based on empirical data once again, but using the area-to-mass ratio as the independent variable.  A delta-$v$ distribution is generated for each value of area-to-mass ratio, which is randomly sampled to generate an ejection speed for each debris fragment.

In summary, our code implements the equations in \cite{2001AdSpR..28.1377J} and proceeds by supplying the model with a value for $M$ and a lower limit for $L_{c}$.  The powerlaw distribution for the fragment characteristic length is formed, generating a total of $N$ fragments.  For each characteristic length bin in this distribution, area-to-mass values are sampled from the appropriate distribution for all fragments in that bin.  These area-to-mass values are used to generate an ejection speed for each fragment, sampled from the appropriate delta-$v$ distribution.  We verified the output of our code against the results in \cite{2001AdSpR..28.1377J}, in particular by accurately replicating Figure 5 from that paper.

Thus, a single run of our code generates a fragment list with a mass for each fragment and an ejection speed for each fragment (along with area-to-mass and characteristic length).  Note that these quantities are necessary but not sufficient to describe a debris cloud formed from a collision.  In addition to an ejection speed, each fragment requires an ejection direction, which is not described by the EVOLVE 4.0 model.  Also, certain physical constraints on the collision should be observed, for example the conservation of momentum.

Further, there is nothing in the EVOLVE 4.0 model to ensure that the sum of the fragment masses is equal to the collision mass used as input into the model.  For example, the sampling of the various distributions is not driven by a mass conservation boundary condition.  In addition, as EVOLVE 4.0 is guided by empirical data, for small fragment sizes/masses little or no data exist, meaning that the models are unconstrained in various fragment size regimes.  

We take a practical approach to introducing these considerations into the simulation and addressing these limitations.  We utilise the output of our EVOLVE 4.0 code in a loop, with exit from the loop requiring certain conditions to be met.  Within the loop, we take our fragment list and assign each fragment a random direction.  We can therefore calculate the momentum for each fragment in the centre of mass frame of reference.  If we find that the magnitude of the total momentum vector is more than 1\% of the sum of the momentum vector magnitudes over all fragments ($|{\bf P_{tot}}|>0.01\Sigma_{n=1}^{N}|{\bf P_{n}}|$), a new set of random directions is calculated and the condition re-tested.  While this is an arbitrary condition, when the condition is met the conservation of momentum is approximately achieved.  At the level of realism we require, and is afforded by the EVOLVE 4.0 model in general, this approximation is acceptable in order to generate a realistic debris cloud in a practical fashion.

In terms of mass conservation, we have investigated the mass budget of the fragments in different collision regimes and have selected realisations of the models for which the total fragment mass is plausibly close to the collision mass input into the model, described further in the following section.

As our code is practically configured, it is fast to run and allows the resulting dynamic debris clouds to be quickly realised at different altitudes, over different collision masses, and for different orbit orientations.  The clouds can be evolved in the gravitational field of the Earth.  In all cases, we can examine how the debris cloud relates to the MWA's observational characteristics, in terms of sensitivity to fragment size and field of view coverage.  In the next section, we present some of the results of this examination.

\section{Results}
\label{results}
In order to demonstrate our simulations and illustrate the salient aspects of the MWA's likely performance in detecting and characterising debris clouds, we run our EVOLVE 4.0 model code using collision masses of 100 kg, 500 kg, and 1000 kg.  In all cases, we use a lower limit of $L_{c}>0.115$, guided by \cite{2001AdSpR..28.1377J}.  The outputs of our implementation of EVOLVE 4.0 were treated as described above, to realise three different debris clouds with as close as possible to mass conservation between input mass and total fragment mass.  For the three collision masses, we achieved approximately 66 kg, 312 kg, and 614 kg for the 100 kg, 500 kg, and 1000 kg collision masses, respectively (setting a minimum 60\% threshold for recovery of mass from a collision in our simulation).

\cite{2001AdSpR..28.1377J} provides information on the total mass budget to augment the simulation of the debris cloud based on $L_{c}>0.115$, in terms of expressions for the mass contained in fragments with $L_{c}<0.08$.  By running 30,000 realisations of these expressions, we find a maximum mass in $L_{c}<0.08$ fragments of approximately 39 kg, 110 kg, and 175 kg for 100 kg, 500 kg, and 1000 kg collisions masses, respectively.  The remaining characteristic length range of $0.08<L_{c}<0.115$ is covered in \cite{2001AdSpR..28.1377J} by reference to a ``bridging function'', which is not described in detail in the paper\footnote{We also enquired directly with the authors regarding the bridging function, but did not receive a response to our enquiry.}

Given the masses above, in the $0.08<L_{c}<0.115$ range, no additional fragments are required for the 100 kg collision mass (because we have evaluated and summed the maxima from repeated realisations of two independently calculated distributions).  For the 500 kg collision mass, we require 78 kg of fragments in the $0.08<L_{c}<0.115$ range.  And for the 1000 kg collision mass, we require 211 kg in this characteristic length range.  Without the detail of the bridging function, we can only assert that these numbers are plausible and that we can plausibly account for the total collision masses.  As noted earlier, we only simulate the dynamics of the collision for characteristic lengths of $L_{c}>0.115$, as this range in the model is constrained by observational data.  

The parameter distributions and results we present beyond this point all refer to the simulation outputs for characteristic lengths of $L_{c}>0.115$, where the model of \cite{2001AdSpR..28.1377J} is well constrained by observations.

\begin{figure}[h]
\includegraphics[width=0.5\textwidth]{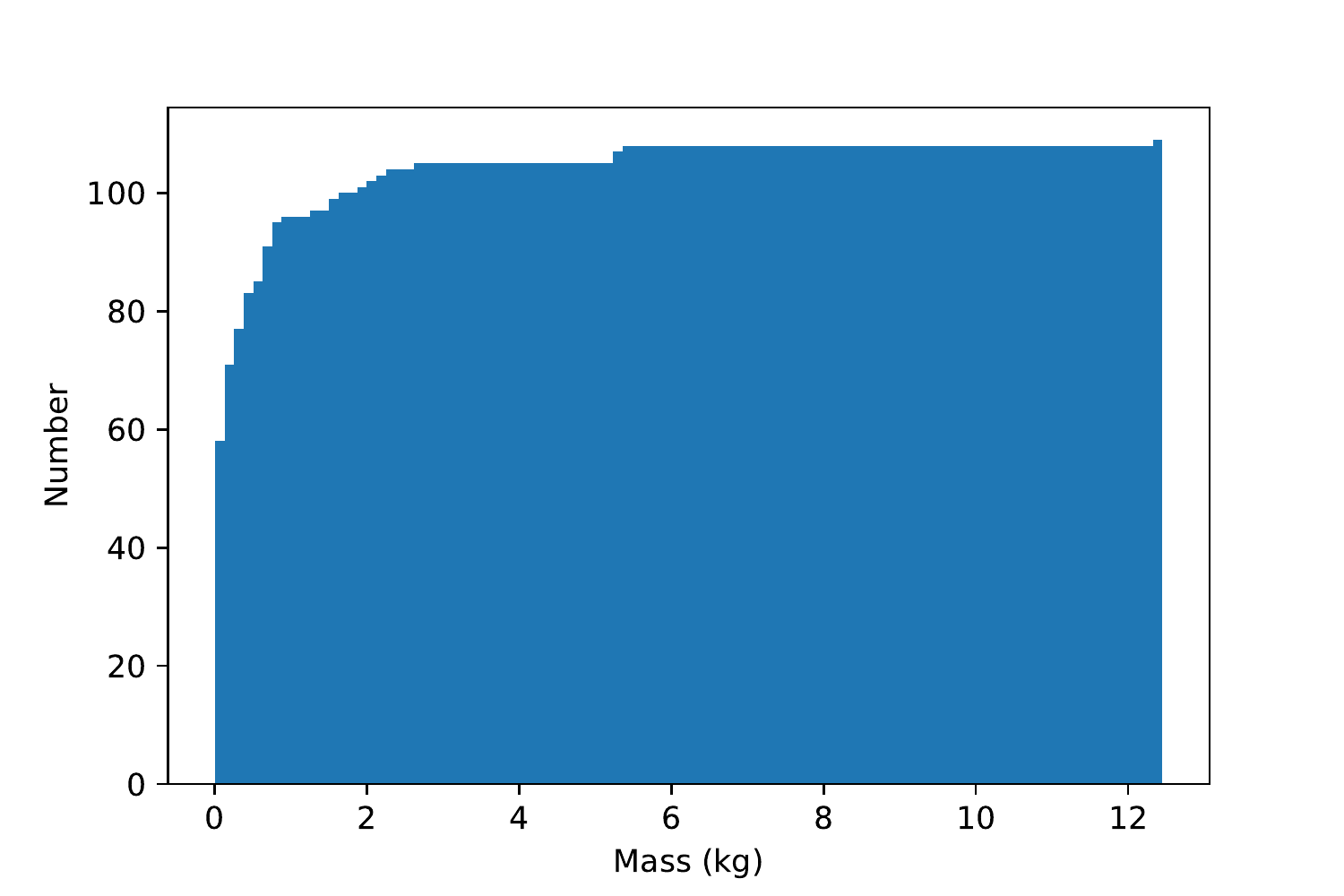}
\includegraphics[width=0.5\textwidth]{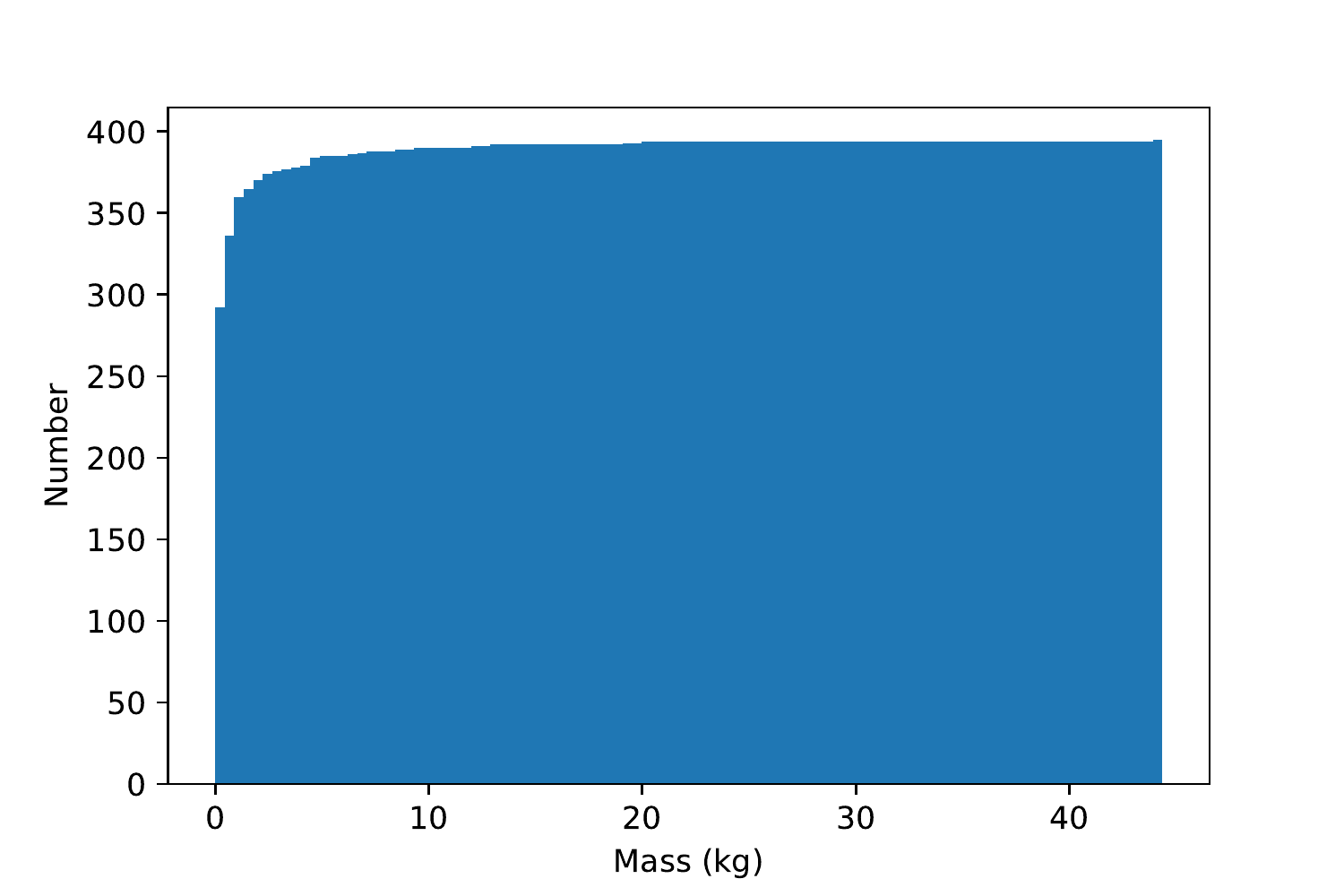}
\includegraphics[width=0.5\textwidth]{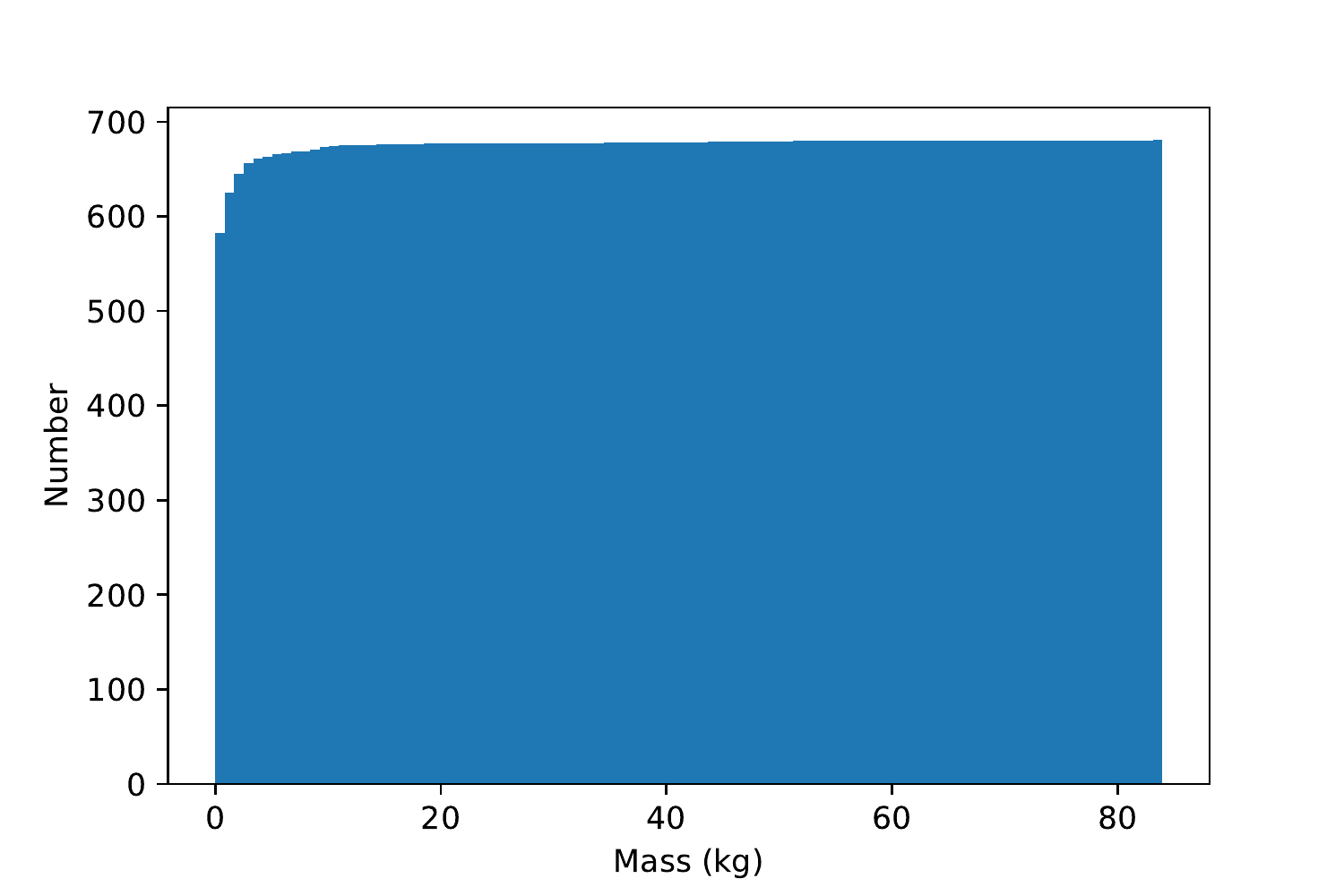}
\caption{Cumulative distributions of fragment masses for 100 kg (top left), 500 kg (top right), and 1000 kg (bottom left) collisions, respectively.}
\label{Fig1}
\end{figure}

\begin{figure}[h]
\includegraphics[width=0.5\textwidth]{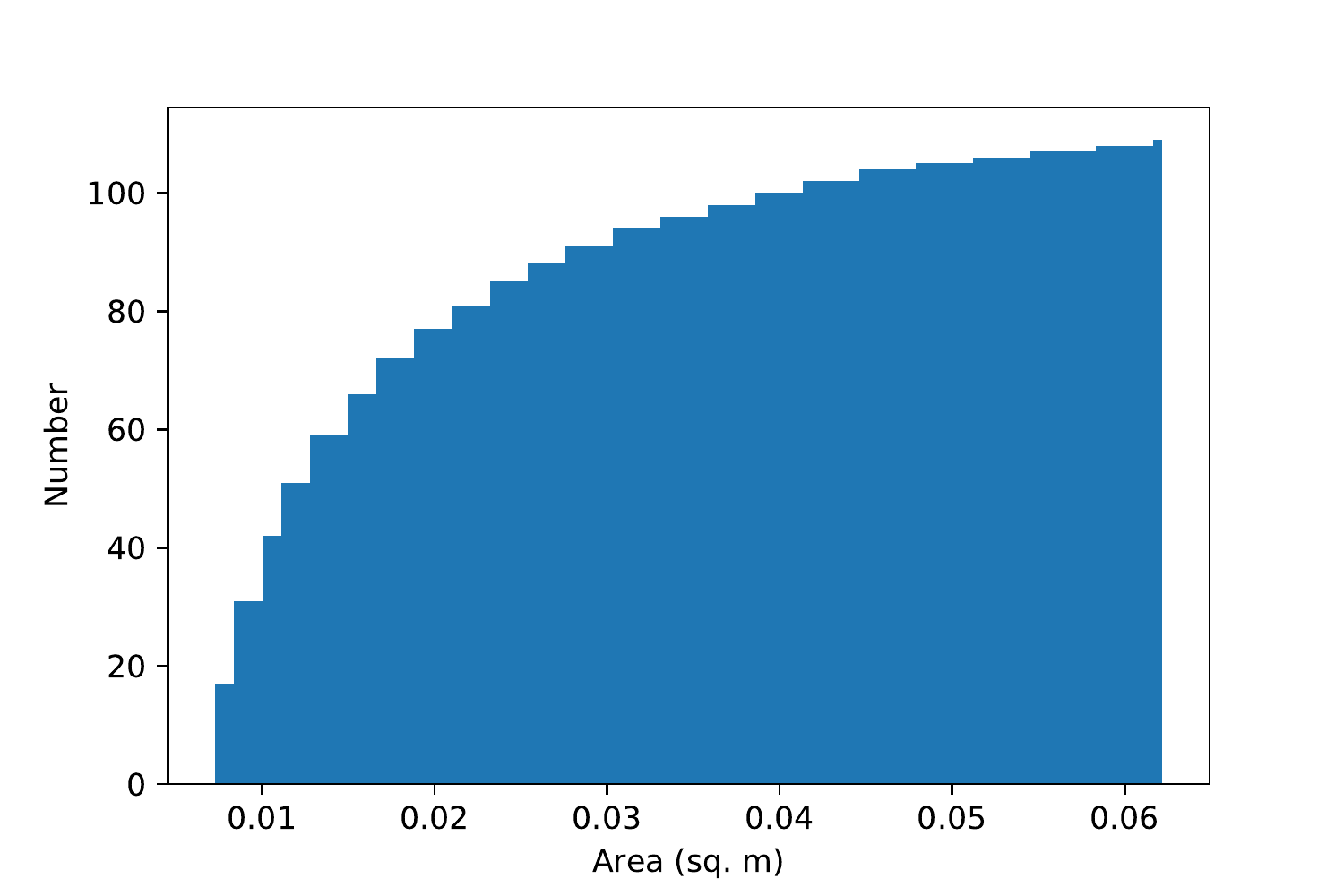}
\includegraphics[width=0.5\textwidth]{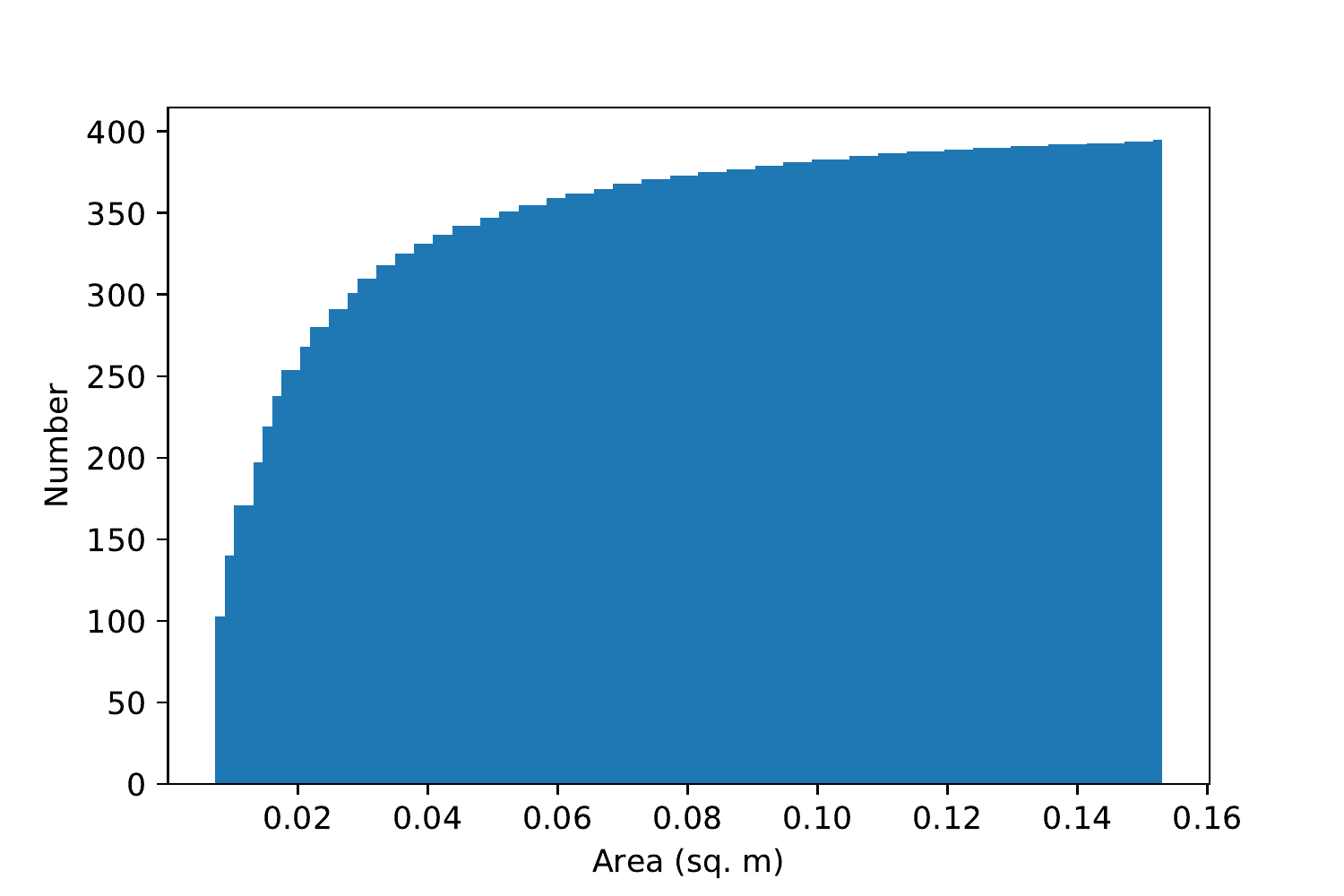}
\includegraphics[width=0.5\textwidth]{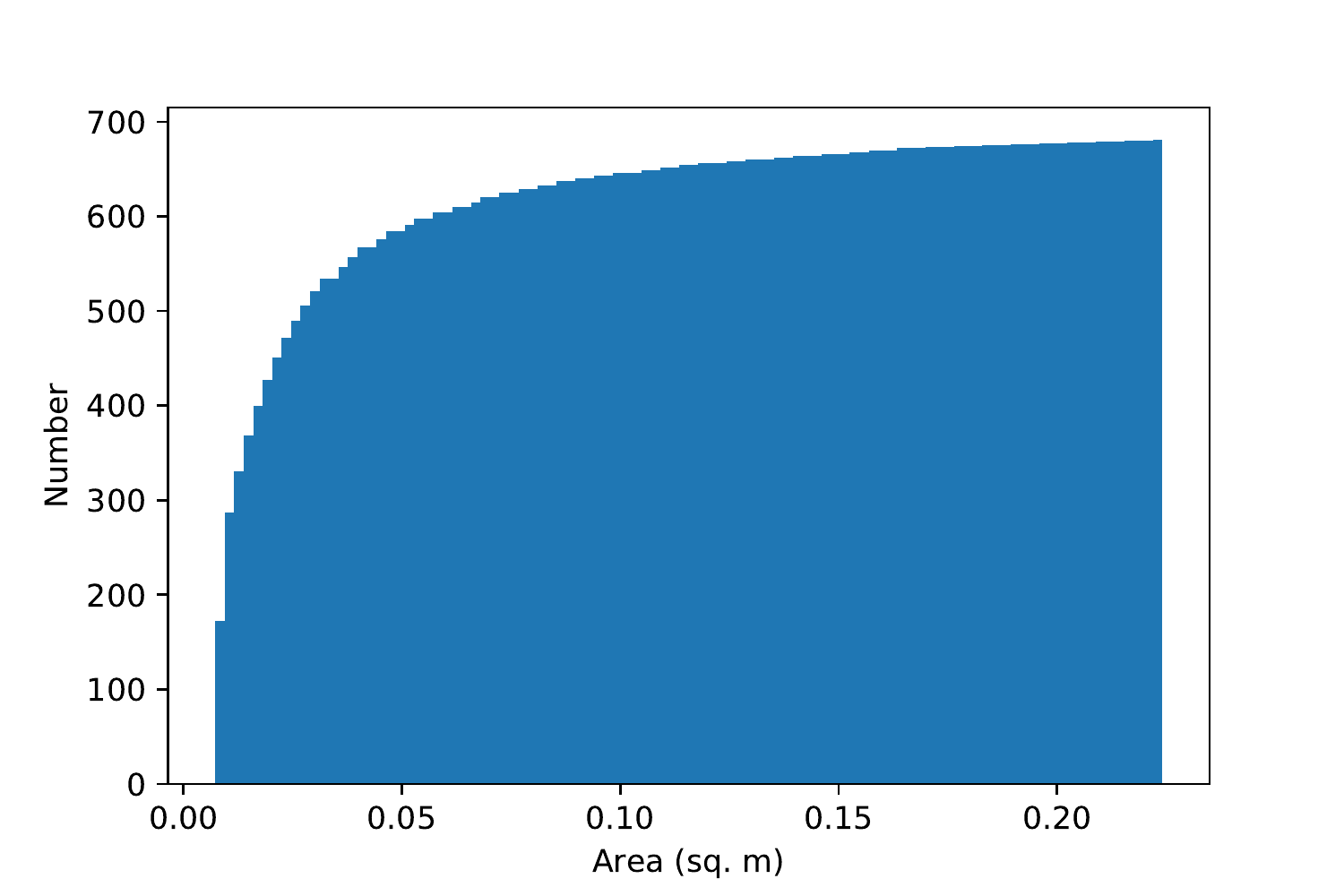}
\caption{Cumulative distributions of fragment areas for 100 kg (top left), 500 kg (top right), and 1000 kg (bottom left) collisions, respectively.}
\label{Fig2}
\end{figure}

\begin{figure}[h]
\includegraphics[width=0.5\textwidth]{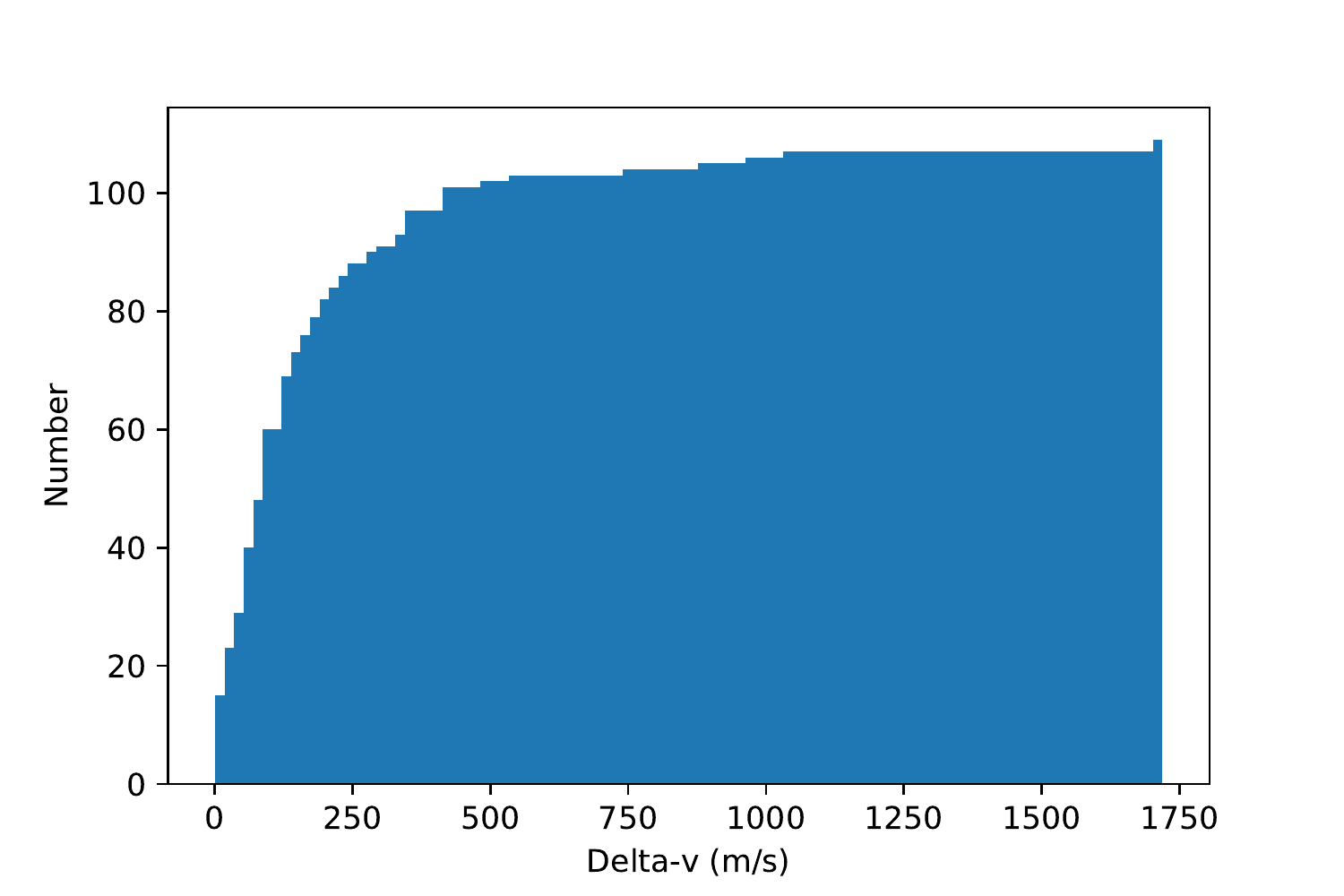}
\includegraphics[width=0.5\textwidth]{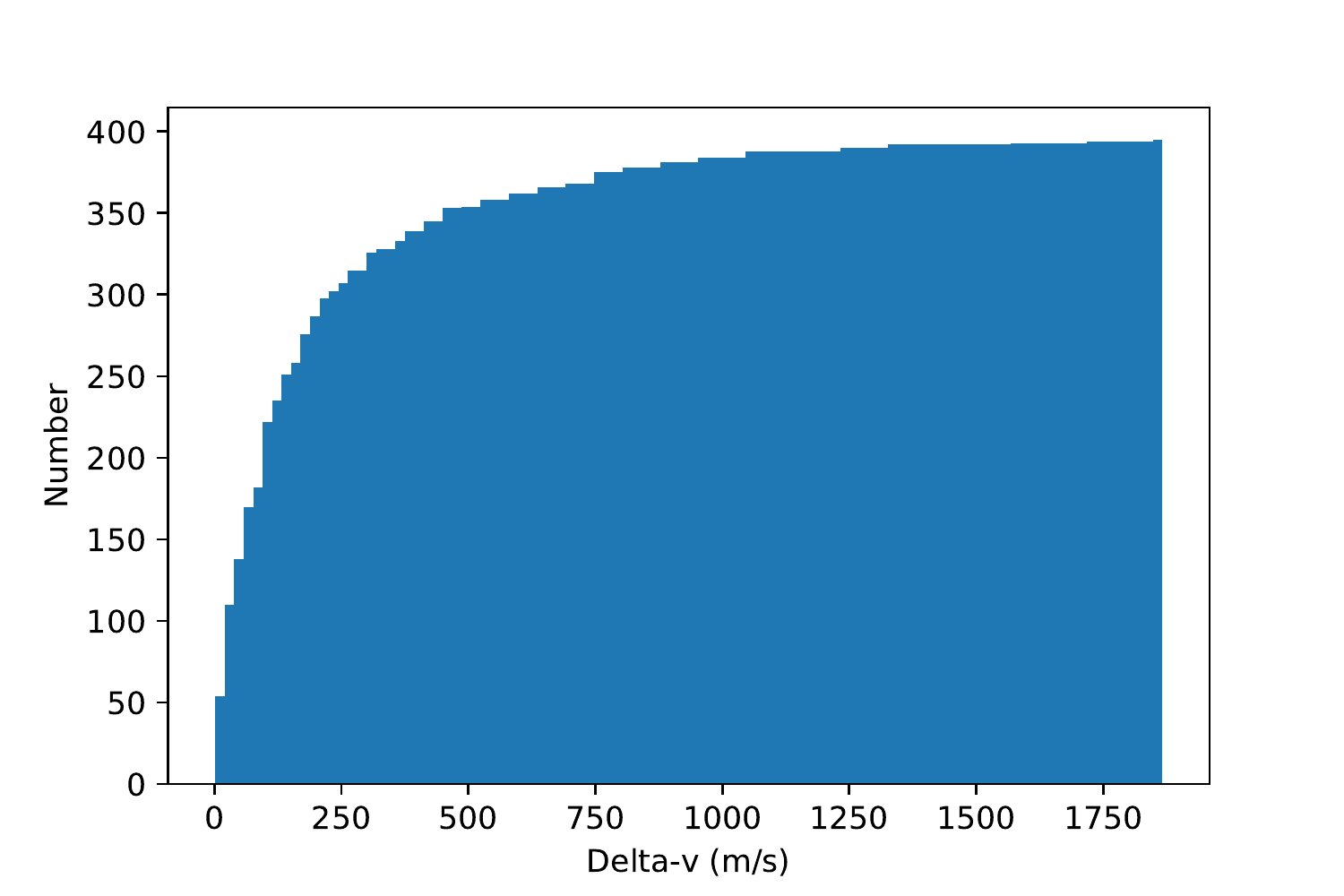}
\includegraphics[width=0.5\textwidth]{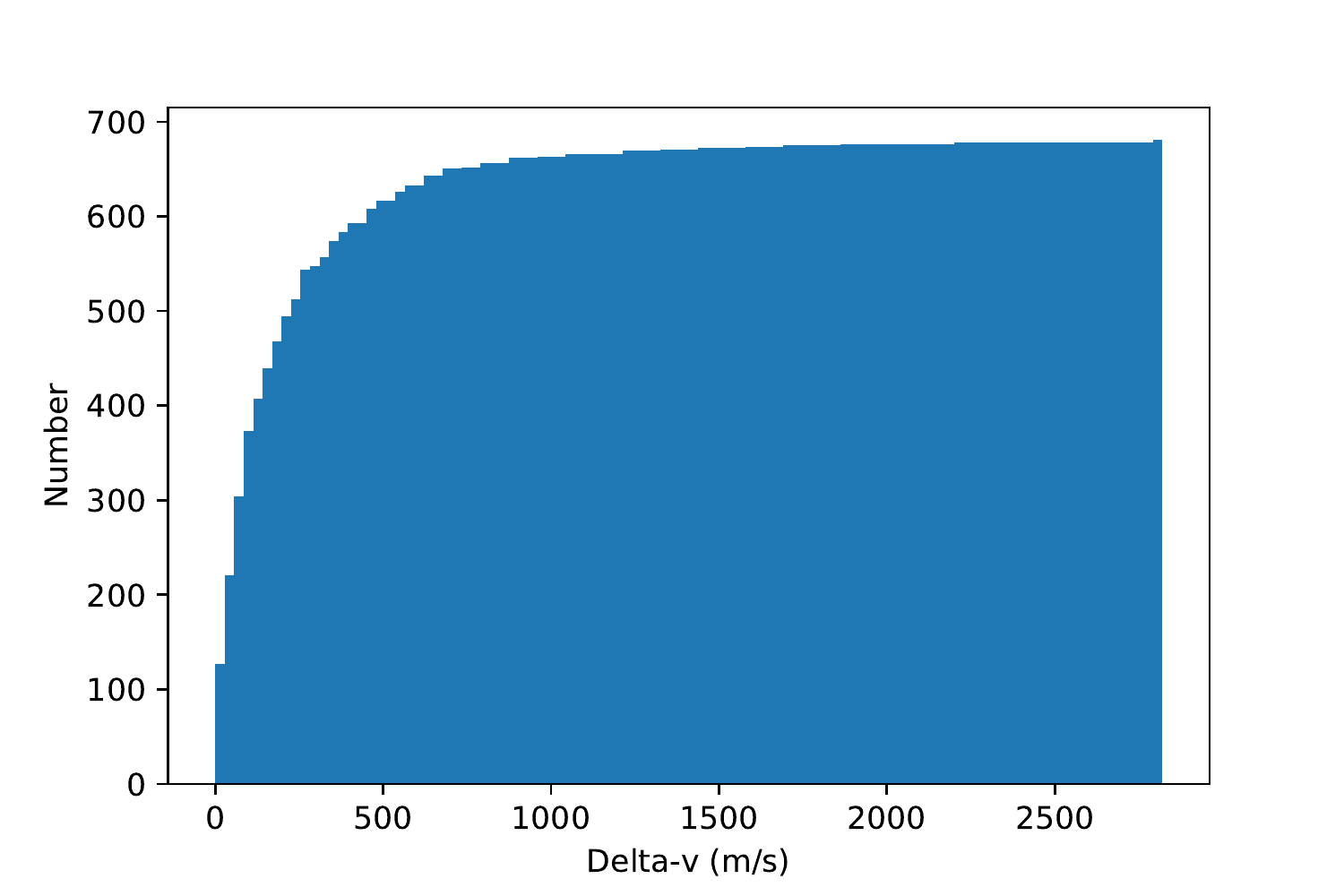}
\caption{Cumulative distributions of fragment delta-$v$ for 100 kg (top left), 500 kg (top right), and 1000 kg (bottom left) collisions, respectively.}
\label{Fig3}
\end{figure}

Figure \ref{Fig1} shows the cumulative distributions of fragment mass for the three collision masses ($L_{c}>0.115$).  Figure \ref{Fig2} shows the cumulative distributions of fragment areas for the three collision masses.  Figure \ref{Fig3} shows the fragment delta-$v$ distributions for the three collision masses.

Further, we locate the three collisions at three different altitudes, 300 km, 800 km, and 1300 km, giving nine different collision scenarios.

\begin{figure}[h]
\includegraphics[width=0.5\textwidth]{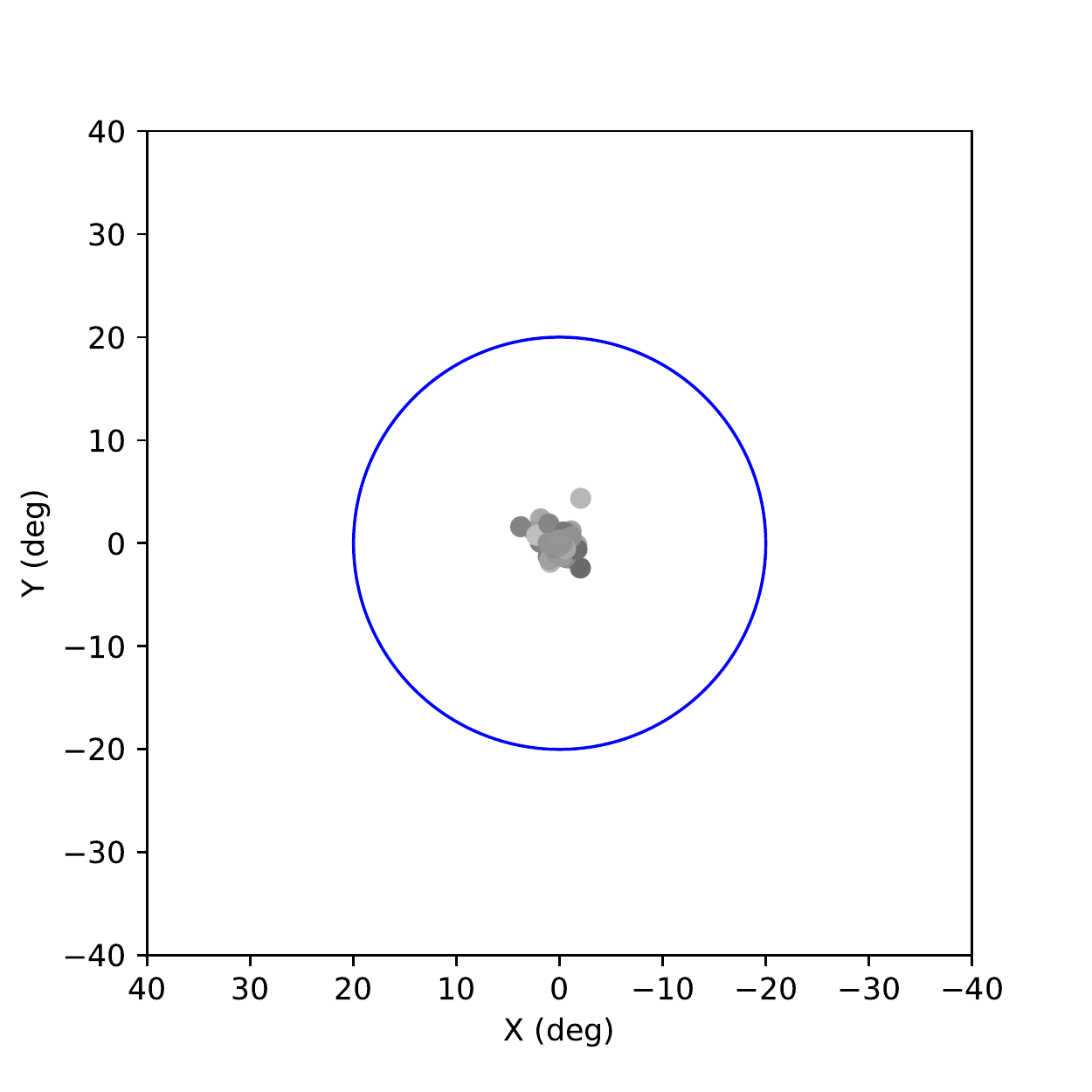}
\includegraphics[width=0.5\textwidth]{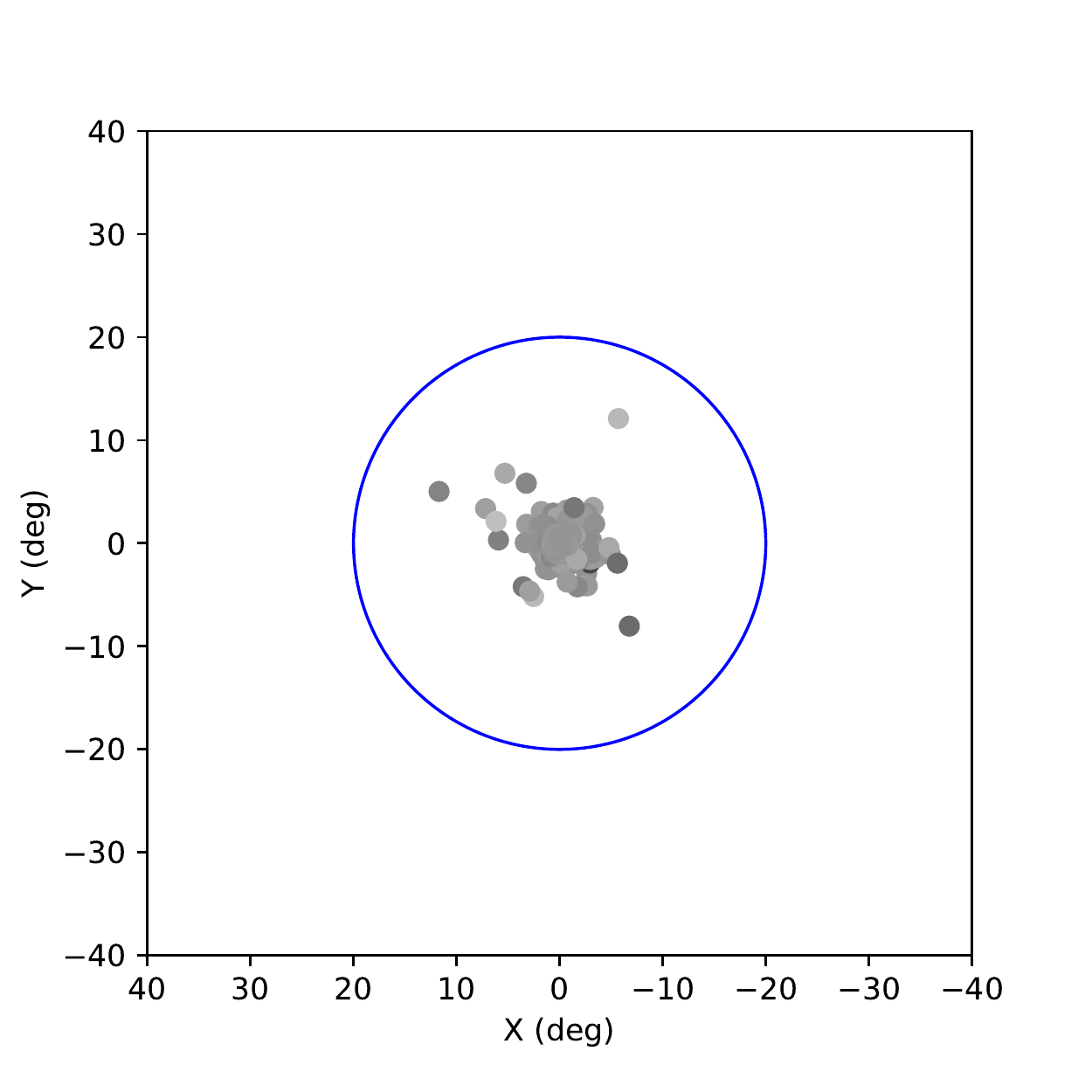}
\includegraphics[width=0.5\textwidth]{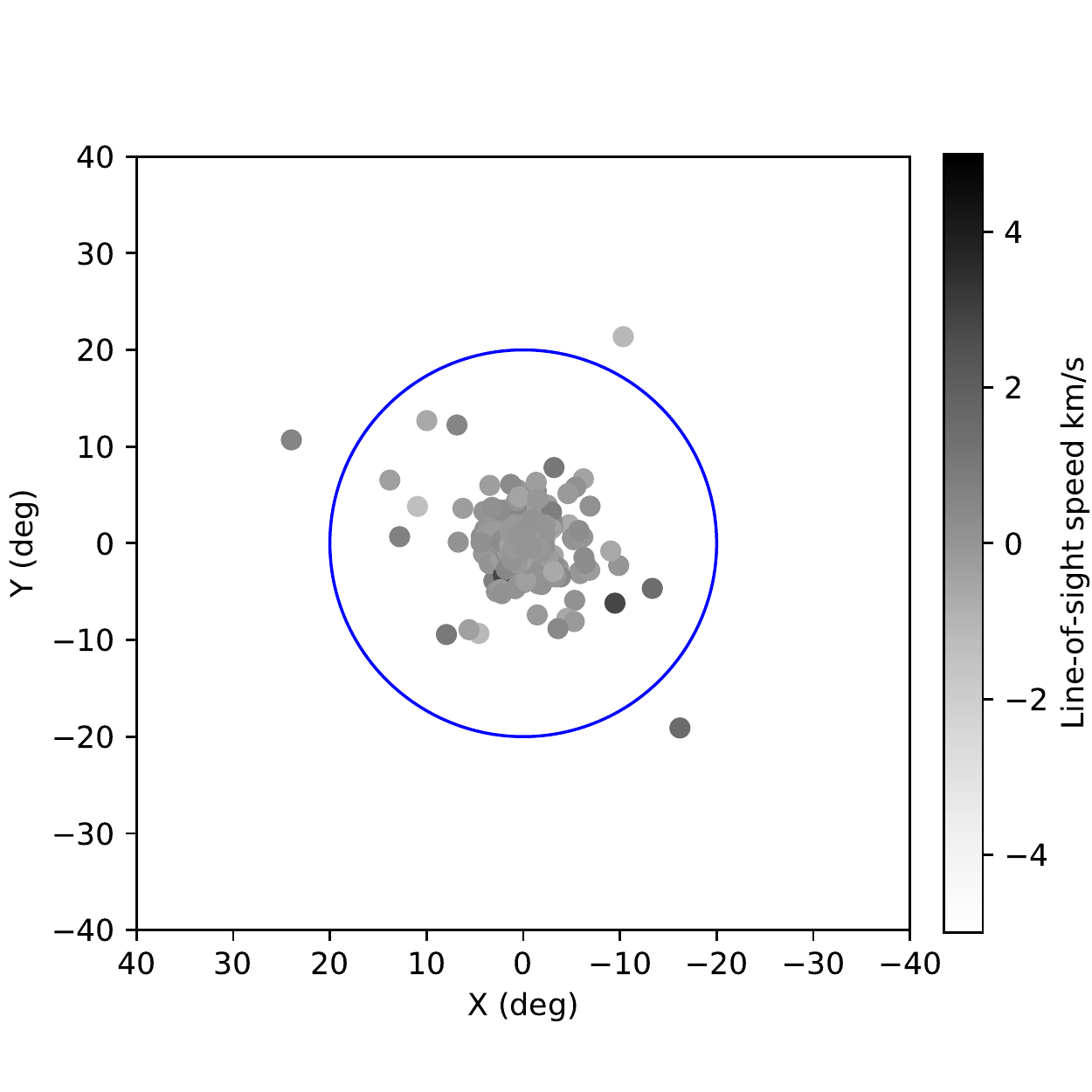}
\caption{Fragment cloud as seen from the MWA, for a collision mass of 1000 kg at an altitude of 300 km, over three time steps at $t=10$ s (top left), $t=30$ s (top right), and $t=60$ s (bottom left).  The circle represents the full width at half maximum (FWHM) field of view on the sky (field of view over which the sensitivity is at least half of the peak sensitivity, 40$^{\circ}$ at 100 MHz observation frequency).  The different shades for the individual fragments denote the line of sight speed of the fragment relative to the MWA, according to the colour bar.}
\label{Fig4}
\end{figure}

For simplicity of representation to the reader, and for the purposes of illustration, we assume that the debris cloud is created in a collision that is close to the zenith for the MWA and that the MWA tracks its pointing direction toward the centre of mass of the collision as the cloud evolves.  Thus, the example in Figure \ref{Fig4} represents the MWA field of view co-moving with the debris cloud across the sky, at three different time steps. 

We recognize that this scenario, while convenient for illustration, is also optimistic.  It is highly unlikely that a collision will occur at zenith for the MWA.  Overwhelmingly, for a collision that occurs above the MWA's local horizon, the range to the event will be greater than if it occured at zenith.  For example, at an altitude of 300 km and a zenith angle of 45$^{\circ}$, the range is 415 km.  These situations are handled easily by our code, but are not shown, for simplicity.  Even more likely is that collisions occur below the MWA horizon and the debris cloud is not visible to the MWA for multiple orbits.  In these cases, the MWA's wide field of view advantages remain, however are reflective of the MWA's general existing utility as a passive radar facility.  The characterization of highly evolved debris clouds is beyond the scope of this paper.

It should also be noted that, even though the following simulated examples observe the debris cloud from its origin time, realistic scenarios will require some time for the MWA to acquire the cloud as a target.  Once a collision is understood to happen and this is communicated to the MWA, the re-pointing response time is approximately 10 seconds to acquisition.  However, in many cases, possible collisions will be predicted in advance (as for the IRAS/GGSE 4 near miss), allowing observations to be optimally scheduled.

\begin{figure}[h]
\begin{centering}
\includegraphics[width=0.8\textwidth]{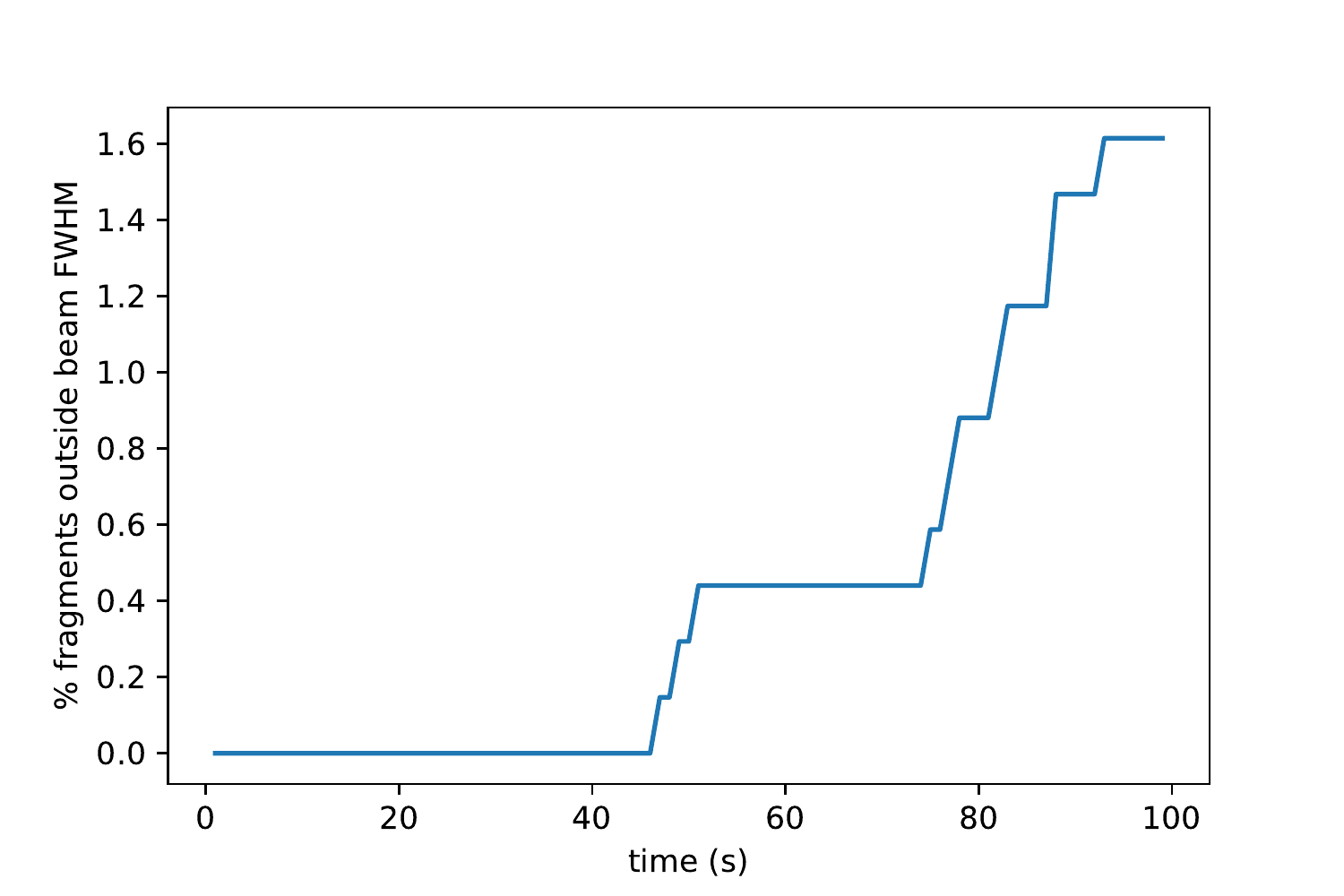}
\caption{The percentage of the debris fragments outside the sensitive portion of the MWA field of view, as a function of time, for a 1000 kg collison at an altitude of 300 km.  The MWA beam FWHM is calculated for an observation frequency of 100 MHz and is approximately 40$^{\circ}$.}
\label{Fig5}
\end{centering}
\end{figure}

As can be seen from Figure \ref{Fig4}, the complete debris cloud is contained within the sensitive portion of the MWA field of view for almost 60 seconds, for a 1000 kg collision at an altitude of 300 km (assuming an observation frequency of 100 MHz).  The percentage of the debris fragments outside the sensitive portion of the MWA field of view, as a function of time, is presented in Figure \ref{Fig5}, showing that 99\% of the debris cloud is contained within the MWA field-of-view for $\sim$80 seconds. 

The other important aspect to consider is the MWA detectability threshold.  Assessing the MWA's sensitivity to detect debris is complex, depending on the details of the uncooperative transmitters participating in the passive radar system, the detailed radar cross-section (RCS) of the object, the range to the object, the bandwidth of the radar signal, and the motion of the object (its orbital parameters determining the integration time over which data can be coherently processed).  The MWA will preferentially detect larger objects at shorter ranges, with sensitivity to a given RCS dropping off with larger ranges.  In order to counteract this effect, and detect objects as small as possible at ranges as large as possible, sensitivity can be increased by increasing the Coherent Processing Interval (CPI) used in the passive radar data processing.  Such techniques are currently an area of active study for the MWA, with some initial results presented by \cite{8835821}.  

A single number describing the MWA's sensitivity is therefore not possible to quote.  In non-coherent passive radar mode, \cite{2013AJ....146..103T} found that in a 50 kHz bandwidth and a 1 s processing interval, objects with an area of $\sim$0.8 m$^{2}$ could be detected at ranges up to approximately 1000 km and that objects with an area of $\sim$0.1m$^{2}$ could be detected at ranges up to approximately 300 km.

Traditional coherent passive radar, as demonstrated for the MWA by \cite{8835821}, is significantly more sensitive than the non-coherent passive radar techniques used for the estimates above.  While the limits of coherent passive radar have not yet been reached for the MWA, early indications are that a factor of ten reduction in detectable fragment area, relative to the non-coherent method estimates above, may not be unreasonable.

Thus, to explore the detectability of the fragments in the collision scenarios described here, we consider fragment area cutoffs for detection of 0.01 m$^{2}$ and 0.1 m$^{2}$ (optimistic to conservative) for the debris cloud at 300 km, 0.05 m$^{2}$ and 0.5 m$^{2}$ for the debris cloud at 800 km, and 0.1 m$^{2}$ and 1 m$^{2}$ for the debris cloud at 1300 km.

\begin{figure}[h]
\includegraphics[width=0.5\textwidth]{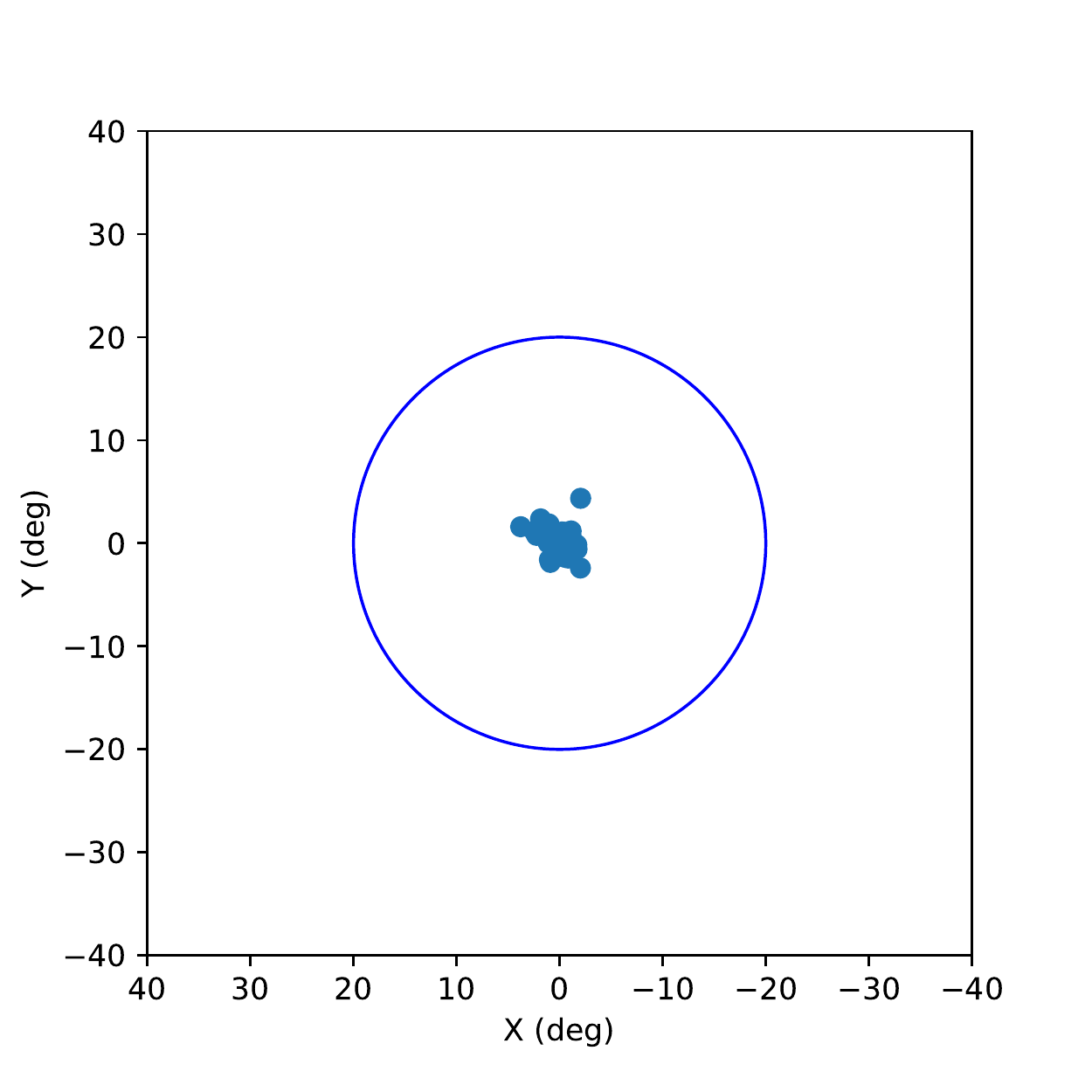}
\includegraphics[width=0.5\textwidth]{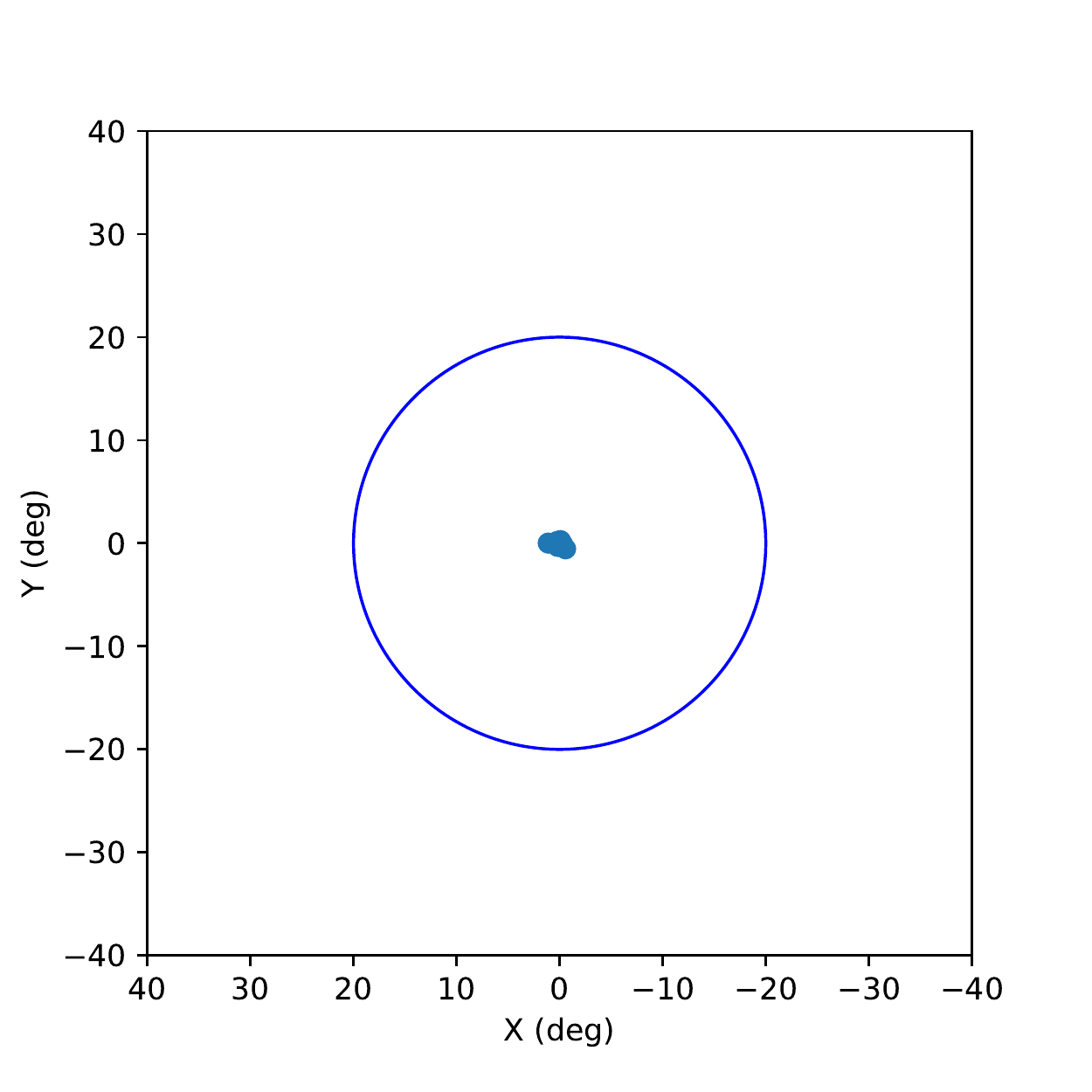}
\includegraphics[width=0.5\textwidth]{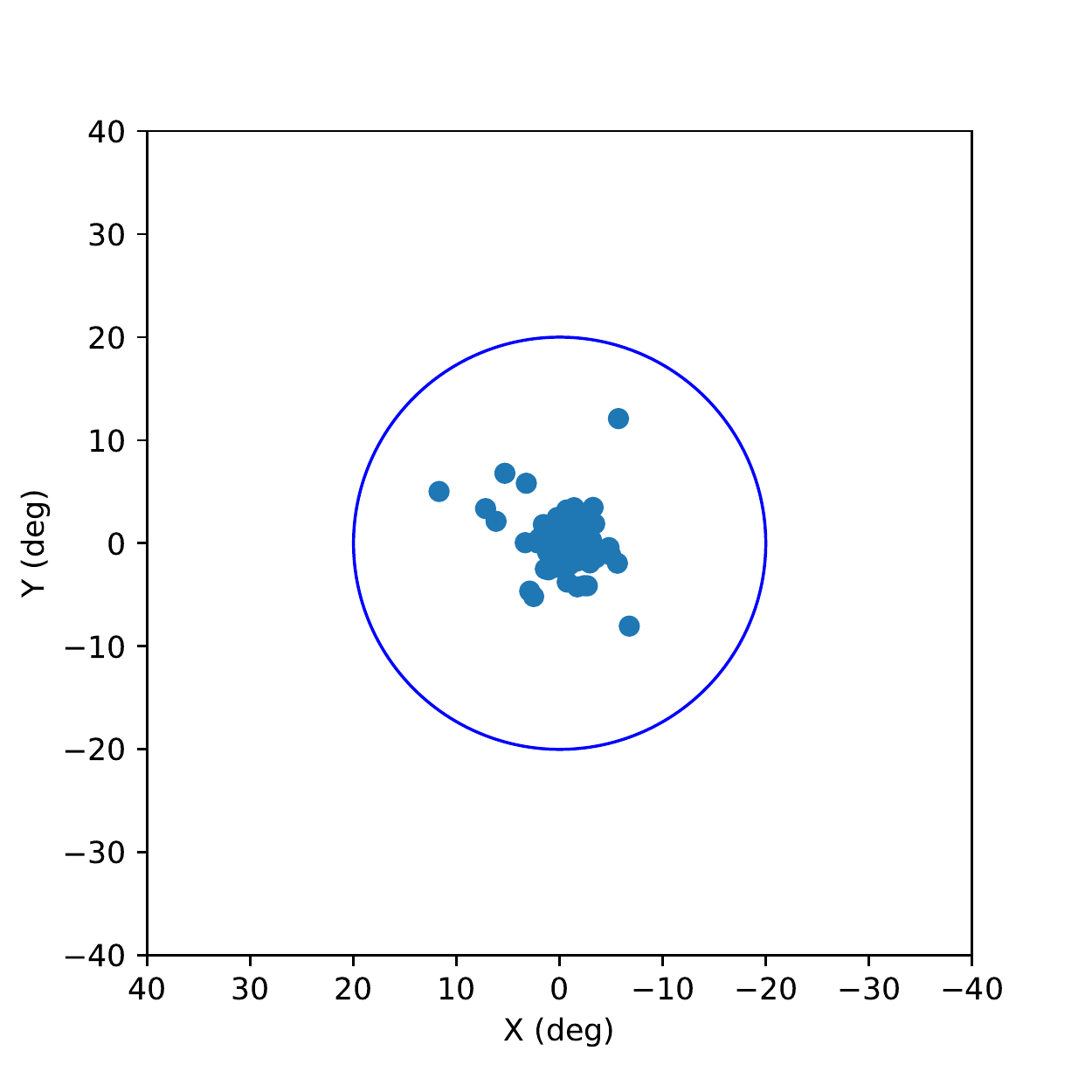}
\includegraphics[width=0.5\textwidth]{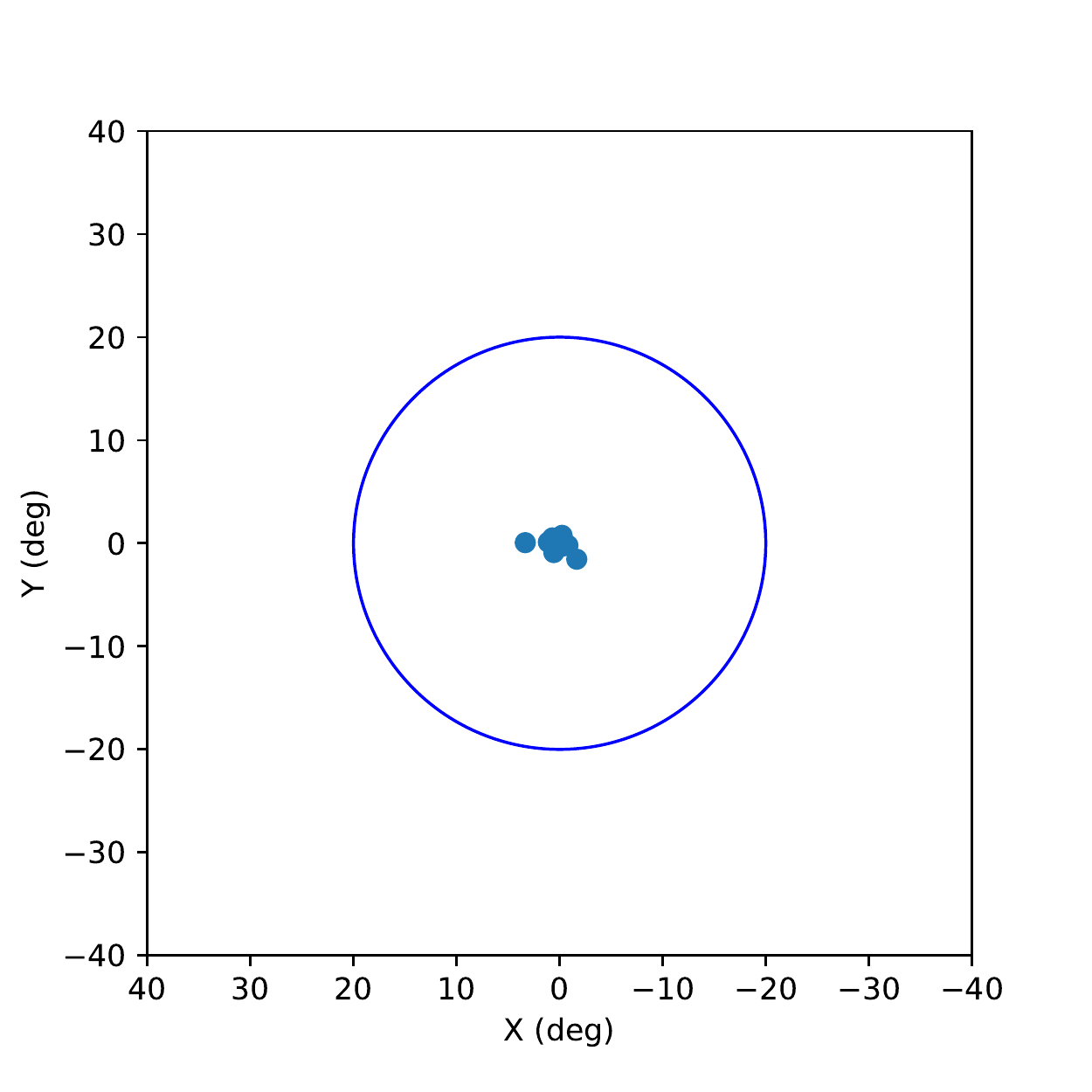}
\includegraphics[width=0.5\textwidth]{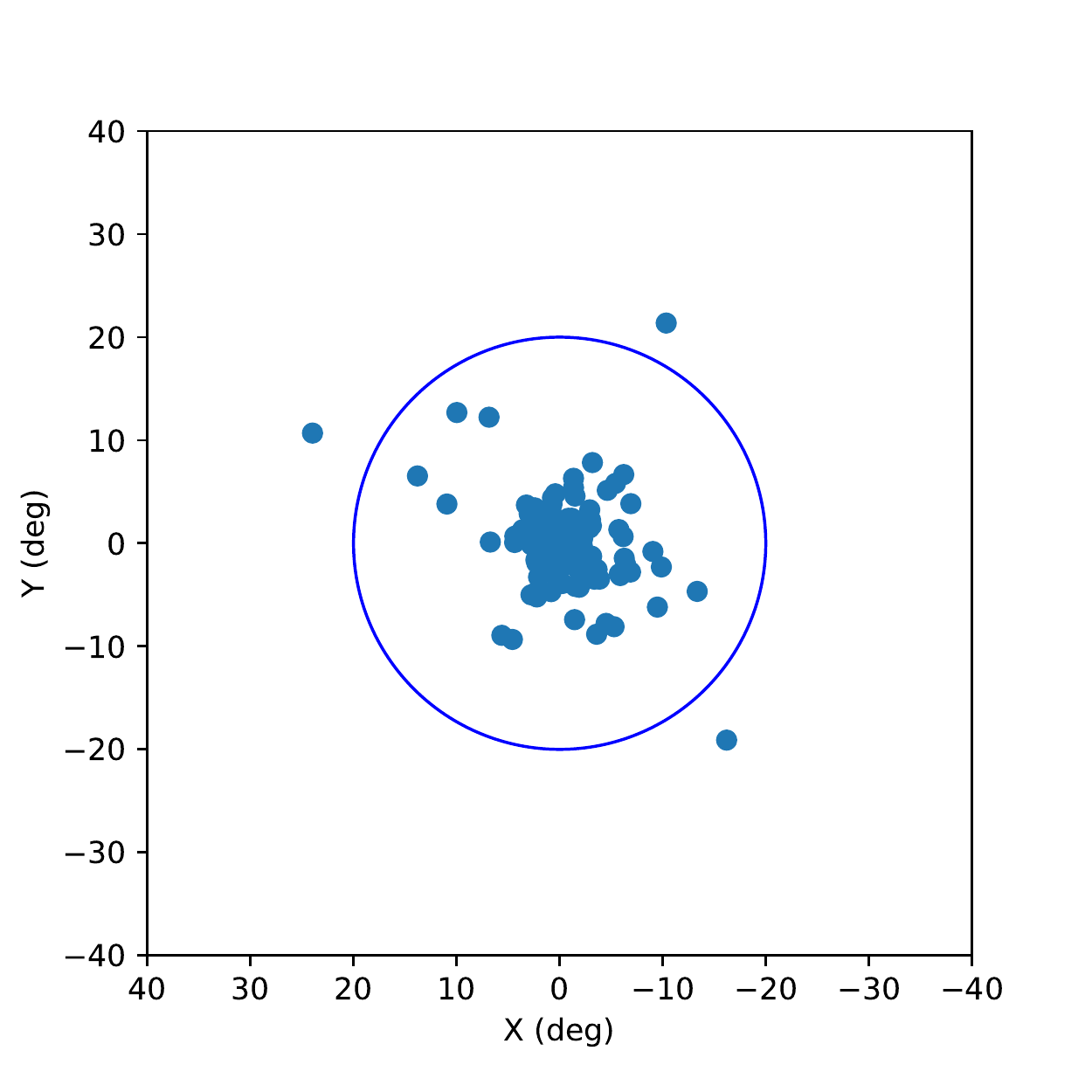}
\includegraphics[width=0.5\textwidth]{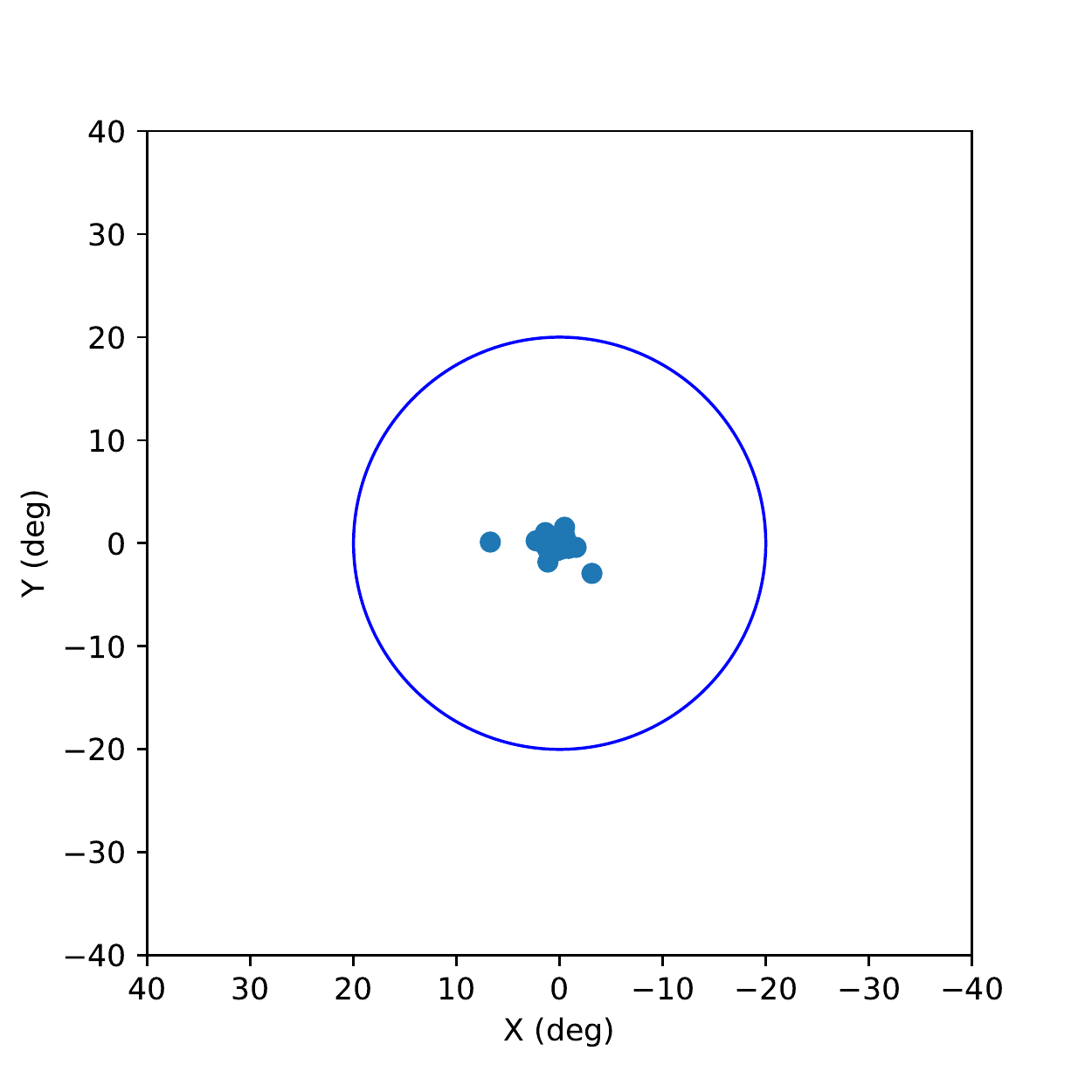}
\caption{Fragment cloud as seen from the MWA, for a collision mass of 1000 kg at an altitude of 300 km, as per Figure \ref{Fig5}, but restricted to objects with area$>$0.01m$^{2}$ (left panels) and area$>$0.1m$^{2}$ (right panels).  Line-of-sight velocity information not included.}
\label{Fig6}
\end{figure}

Figure \ref{Fig6} shows the same as in Figure \ref{Fig4}, but with the fragments plotted restricted to area$>$0.01 m$^{2}$ and area$>$0.1m$^{2}$, to indicate which fragments within the cloud are likely detected for a 1000 kg collision mass at 300 km.  A comparison between the left and right panels of Figure \ref{Fig6} shows that as sensitivity decreases, the smaller and faster fragments preferentially avoid detection.  A summary of the results is provided in Table \ref{tab1}, where the percentage of detectable fragments is listed as a function of collision mass and altitude, for both optimistic and conservative scenarios.

\begin{table}[h]
\begin{centering}
\begin{tabular}{|c|ccc|ccc|} \hline
&\multicolumn{3}{c|}{Optimistic detection scenario}&\multicolumn{3}{c|}{Conservative detection scenario} \\
Collision mass&300 km&800 km&1300 km&300 km&800 km&1300 km \\ \hline
100 kg&72\%&4\%&0\%&0\%&0\%&0\% \\ \hline
500 kg&74\%&12\%&4\%&4\%&0\%&0\% \\ \hline
1000 kg&75\%&14\%&6\%&6\%&0\%&0\% \\ \hline
\end{tabular}
\caption{Percentage of detectable fragments as a function of collision mass and altitude, in optimistic (left) and conservative (right) scenarios.  Collision masses in left hand column.  Altitudes in top row.}
\label{tab1}
\end{centering}
\end{table}

\section{Discussion and Conclusions}
\label{dandc}

Our results show that a high percentage ($>70\%$) of fragments generated in the simulated debris clouds with $L_{c}>0.115$ are detectable in low altitude orbits (300 km), for optimistic detection scenarios, for all collision masses.  At higher altitude orbits (or greater ranges), the detection fraction drops off to $\sim10\%$ at 800 km and below 10\% at 1300 km.  

For conservative detection scenarios (fragment area greater than 0.1 m$^{2}$), large collision masses ($>500$ kg) generate a small but significant number of detectable fragments ($\sim$5\%) in low orbits ($\sim$300 km).  But small collision masses in low orbits do not produce fragments of a detectable size in the conservative scenario, which is likewise true for larger collision masses in higher orbits (refer to Table \ref{tab1}).

While these results cannot be considered definitive for all collisions, such as glancing collisions, we have assumed catastrophic collisions (and have modeled as such from EVOLVE 4.0) and the results do point to general conclusions regarding the detection performance of the MWA for debris clouds.  Detection is clearly easier for large numbers of fragments for large collision masses in low orbits and, in general, the MWA will have to realise close to the optimistic detection scenario to be maximally useful.  Thus, current efforts to improve our detection limits \cite{8835821} should continue, in an attempt to reach or surpass the optimistic scenario explored here.  However, even in the case of the conservative scenario, useful numbers of fragments are detectable by the MWA in the lower and most populated orbits of LEO.  So, even the minimal expectations on MWA detection performance are useful, since even the minimal expectations are adequate to determine whether or not a collision has occurred.

The second area of MWA performance examined in this work is the extent to which the evolving debris cloud is contained within the MWA field of view.  We find extremely favourable results in this measure, with all simulations across all collision masses and altitudes showing that at least 98\% of the debris fragments are contained within the sensitive portion of the MWA field of view for at least 100 seconds.  This means that with a single observation of duration less than 100 seconds, all detectable fragments can be detected and their trajectories characterised, in the low probability case that a collision occurs above the MWA local horizon.  However, even if acquisition of the cloud occurs multiple orbits after the collision, the wide-field nature of the MWA will allow the cloud to be comprehensively covered rapidly, in a small number of observations.

Overall, then, in the passive radar observation mode, the MWA can collect a single (or small number of) $\sim$100 second dataset(s) that contain all information on fragments reaching the MWA detection limits.  The final consideration, then, is how quickly these 100 seconds of data can be processed in order to characterise the fragment trajectories and issue that information to interested parties. Currently, as passive radar techniques for the MWA are under active research development, data are collected and stored for processing at a later date, as described in \cite{2013AJ....146..103T,7944483,8835821}. Generally, this processing utilises 10s of seconds worth of data to achieve detection to the MWA sensitivity limits.  With present techniques and implementations, the observation, data processing, and characterisation of a debris cloud could optimistically be achieved in $\sim10$ hours (several orbital periods in LEO).  Thus, in theory the intrinsic capability of the MWA to observe a full debris cloud and attempt detection across all fragments is evident.  In practise, the limit in characterising the debris cloud and provided useful information in a timely manner is limited by the current non-real-time data processing.  This can be solved by an appropriate real-time implementation of the processing codes, more efficient uncued search algorithms (for example already under development by \cite{2020arXiv200303947H}), and computing infrastructure of an appropriate scale.

The results of the work presented here are sufficiently encouraging that steps in this direction are warranted, in order to develop an operational real-time and rapid reaction capability for SSA, with a niche capability for breakup events.

More broadly, the work presented here could be adapted and applied to any SSA sensor type, in order to asses performance for the same task, under whatever range of specific conditions apply.  In the sense of astronomical facilities making an appropriate contribution to global efforts in SSA, the MWA is well-placed to play its part.

\section*{Acknowlegdements}
This research has made use of NASA's Astrophysics Data System.  We have made use of NumPy \cite{van2011numpy}, matplotlib (a Python library for publication quality graphics \cite{Hunter:2007}), SciPy \cite{jones_scipy_2001}, and Astropy \cite{2013A&A...558A..33A} (a community-developed core Python package for Astronomy).  We thank the anonymous referee for comments that improved the paper.


%
%

\bibliographystyle{spmpsci}      

\bibliography{biblio2.bib}

\end{document}